\documentstyle[aps,prl,amstex,epsfig]{revtex} 
\begin{document} 
\draft 
\title{Dynamic Scaling of Magnetic Flux Noise Near the KTB Transition 
in Overdamped Josephson Junction Arrays} 
\author{ 
T. J. Shaw,$^{1,2}$ M. J. Ferrari,$^{1,2,}$\cite{MJFaddress} 
L. L. Sohn,$^{3,}$\cite{LLSaddress} D.-H. Lee,$^1$ 
M. Tinkham,$^3$ and John Clarke$^{1,2}$
} 
\address{$^1$Department of Physics, University of California, Berkeley, 
California  94720\\ 
$^2$Materials Sciences Division, Lawrence Berkeley Laboratory, 
University of California, Berkeley, California  94720\\ 
$^3$Department of Physics and Division of Applied Sciences, 
Harvard University, Cambridge, Massachussetts  02138 
} 
\date{\today} 
\maketitle 
\begin{abstract} 
We have used a dc Superconducting QUantum Interference Device to 
measure the magnetic flux noise generated by the equilibrium vortex 
density fluctuations associated with the Kosterlitz-Thouless-Berezinskii 
(KTB) transition in an overdamped Josephson junction array. 
At temperatures slightly above the KTB transition temperature, the 
noise is white for $f<f_\xi$ and scales as $1/f$ for $f>f_\xi$.  Here 
$f_\xi\propto\xi^{-z}$, where $\xi$ is the correlation length and $z$ is 
the dynamic exponent.  Moreover, when all frequencies are scaled by 
$f_\xi$, data for different temperatures and frequencies collapse on to 
a single curve. 
\end{abstract} 

\narrowtext\twocolumn
Arrays of Josephson junctions have been used extensively as a model 
system to study the effects of order parameter phase fluctuations on 
the superconducting transition in two dimensions.  Such arrays can be 
fabricated with a high degree of uniformity and their relevant 
parameters can be accurately determined.  It is widely accepted that 
the zero field transition is described by the Kosterlitz-Thouless- 
Berezinskii (KTB) theory \cite{kosterlitz73,kosterlitz74,berezinskii70,%
lobb83} and its extension to non-zero frequency \cite{halperin79,%
ambegaokar80,shenoy85}.  According to this theory, phase coherence 
is established throughout the sample below a 
temperature $T_{\rm{}KTB}$, and the system is superconducting.  For 
temperatures above $T_{\rm{}KTB}$ but below the bulk 
transition temperature, even though individual islands are 
superconducting the array is not. The thermal excitations, 
vortices and antivortices, that trigger this phase transition are topological 
defects of the order parameter.  Below $T_{\rm{}KTB}$ vortices and 
antivortices bind in pairs to produce a vortex dielectric, while above 
$T_{\rm{}KTB}$ the pairs dissociate to form a vortex plasma.  In the 
vortex plasma phase, one can identify a characteristic length, $\xi$, as 
the average separation between free vortices;  as $T\rightarrow 
T^+_{\rm{}KTB}, \xi$ diverges.  Thermal fluctuations that perturb the 
vortex density away from its equilibrium value relax through some local 
dynamic process.  Thus, associated with the characteristic length $\xi$ 
there is a characteristic time $\tau$ (or an inverse characteristic 
frequency $f^{-1}_\xi\propto\tau$) corresponding to the time required for 
the disturbance to propagate across the distance $\xi$.  As $\xi$ 
diverges, so does $\tau\ (f_\xi^{-1})$, signifying critical slowing down.  In 
general $\tau=\tau_0(\xi/\xi_0)^z$, where the exponent $z$ depends on 
the dynamics of the relaxation and $\tau_0$ and $\xi_0$ are 
non-universal time and length scales characteristic to the specific 
sample.  For simple diffusion, $z=2$. 
 
Previous experimental studies have involved both electrical resistance 
\cite{resnick81,voss82,abraham82,webb83,kimhi84,brown86,vanwees87,%
carini88,vanderzant91} and two-coil mutual inductance \cite{leemann86} 
techniques, both of which apply an external force to the system and are 
generally confined to a specific frequency.  Because the transition to the 
resistive state is determined by the dissociation of vortex pairs by thermal 
fluctuations, these external forces, which also dissociate pairs, affect 
one's ability to study the intrinsic critical phnomena near 
the true thermodynamic transition temperature. In this paper we employ 
a {\it non-invasive} probe to study the transition in equilibrium.  
Specifically, we measure the spectral density of magnetic flux noise 
\cite{lerch92}, $S_\Phi(f)$, over a frequency range of more than five 
decades.  We find that $S_\Phi(f)$ is white for $f<f_\xi$ and scales as 
$1/f$ for $f>f_\xi$.  In addition, by plotting $fS_\Phi(f)$ versus $f/f_\xi$ 
we show that the data collapse in a manner consistent with dynamic 
scaling. 
 
\begin{figure}
\begin{center}
\epsfig{file=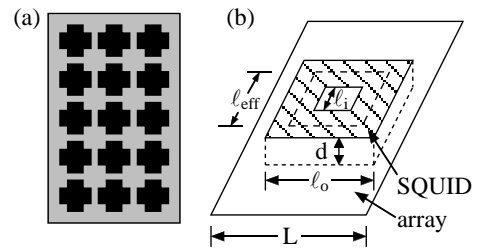,width=3in}
\end{center}
\caption{(a) Schematic of array.  Crosses are niobium islands; area 
between crosses is copper film.  (b) Schematic representation of SQUID, 
with inner and outer dimensions $\ell_i$ and $\ell_o$, a distance $d$ from
array; dashed square has area $\ell_{\rm{}eff}^2=\ell_i\ell_o$.}
\label{schematic}
\end{figure}

The $1\text{ mm}\times{}1\text{ mm}$ array \cite{sohn92} consists of 0.2 
$\mu$m-thick cross-shaped niobium islands patterned on top of a 0.3 
$\mu$m-thick copper film [Fig.\ \ref{schematic}(a)].  The islands form a 
square array with a lattice constant of 10 $\mu$m, and the junctions are 
4 $\mu$m wide and 2 $\mu$m long.  The measurement apparatus, 
originally used to study vortex motion in high-temperature 
superconductors \cite{ferrari94}, involves a Nb-based Superconducting 
QUantum Interference Device (SQUID) attached to a cold stage inside a 
vacuum can surrounded by liquid $^4$He.  A mu-metal shield reduces the 
static magnetic field to less than 1 $\mu$T.  The array, equipped with 
current and voltage leads, is mounted a distance $d\ (<100 \mu\text{m})$ 
away on a variable temperature stage.  The SQUID is a square washer with 
inner and outer dimensions $\ell_i=180\ \mu\text{m}$ and $\ell_o=900 
\ \mu\text{m}$ [Fig.\ \ref{schematic}(b)], and is operated in a flux-locked 
loop.  The output signal is proportional to the change in magnetic flux 
through the SQUID induced by vortex motion in the array. 

\begin{figure}
\begin{center}
\epsfig{file=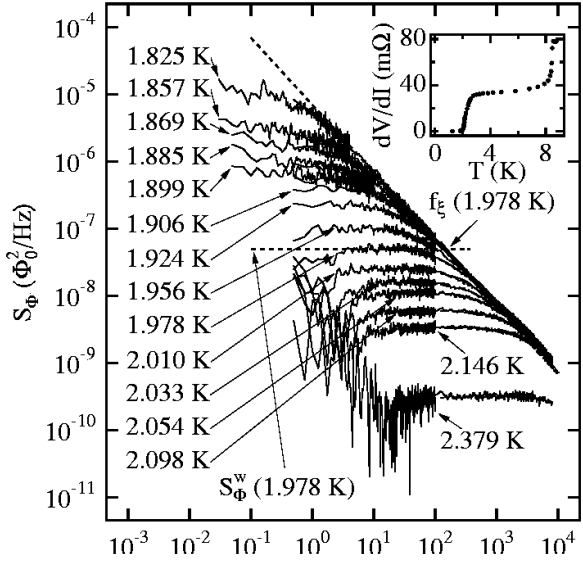,width=3in}
\end{center}
\caption{Spectral density of magnetic flux noise, $S_\Phi(f)$, versus 
frequency for 15 temperatures above $T_{\rm{}KTB}$; scatter at higher 
temperatures is due to subtraction of SQUID noise.  Dashed lines have 
slope $-1$ and 0.  Inset shows $dV/dI$ versus $T$.}
\label{rawdata}
\end{figure}

The inset to Fig.\ \ref{rawdata} shows the differential resistance, $dV/dI$, 
of the array versus temperature, $T$, measured at zero bias current with 
an rms current of 10 $\mu$A at a frequency of 47 Hz.  The initial drop in 
the resistance at approximately $T=8\text{ K}$ marks the bulk transition 
temperature of the niobium islands.  As the temperature is lowered, a 
resistive plateau develops and is followed by a second precipitous drop, 
which is the KTB transition.  These data are similar to those obtained in 
previous experiments \cite{resnick81,voss82,abraham82,webb83,%
kimhi84,brown86,vanwees87,carini88,vanderzant91}.  To obtain the 
average critical current per junction, $i_c(T)$, we measure $dV/dI$ as a 
function of the static bias current.  We take the current at which $dV/dI$ 
is a maximum as $Ni_c(T)$, where $N$ is the number of junctions across 
the width of the array \cite{ambegaokar69,falco74}. 
 
Figure \ref{rawdata} shows the spectral density of flux noise, $S_\Phi(f)$, 
for 15 temperatures above $T_{\rm{}KTB}$.  At each temperature 
$S_\Phi(f)$ is white for $f<f_\xi(T)$ and $\propto{}1/f$ for $f>f_\xi(T)$. 
We define $f_\xi(T)$ as the intersection of lines through the white and 
$1/f$ noise regions as shown in Fig.\ \ref{rawdata} for 
$T=1.978\text{ K}$.  Qualitatively, we understand the difference between 
the two frequency regimes as follows.  For $f>f_\xi$ we probe the system 
at a time scale shorter than that required for vortex density disturbances 
to travel over the correlation length.  Thus, the system appears to be 
critical.  On the other hand, for $f<f_\xi$ the time scale
\noindent 
is longer than the 
characteristic relaxation time, so that the system appears disordered.  In 
this picture, the distinctions between the $1/f$ (critical) and white 
(disordered) power spectra reflect the fundamental differences in the 
relaxation dynamics of these two very different thermodynamic states.  It 
is worthwhile to point out that theories based on the two-dimensional XY 
model and time-dependent-Ginzburg-Landau dynamics predict a $1/f^2$ 
behavior for $f>f_\xi$ \cite{lee,houlrik94}.  This is in marked contrast to 
the $1/f$ behavior observed in our experiment. 

To interpret our data, we now present a brief discussion of a scaling 
theory for the flux noise measurements.  In our geometry 
[Fig.\ \ref{schematic}(b)], $\hat{h}$ is the direction perpendicular to the 
plane $(h=0)$ defined by the array.  We denote a three-dimensional vector 
by $(\vec{x},h)$, where $\vec{x}$ is the component of the vector in the 
plane of the array.  In the absence of the SQUID, the perpendicular 
component of the magnetic field, $B_\perp$, induced at point $(\vec{x},h)$ 
at time $t$ by a vortex density distribution 
$\rho_v(\vec{x}^\prime,t)$ at $h=0$ is 
\begin{equation} 
B_\perp(\vec{x},h;t)=\Phi_0\int{}d^2x^\prime{}K(\vec{x}-\vec{x}^\prime, 
h)\rho_v(\vec{x}^\prime,t). 
\label{bfield} 
\end{equation} 
Here $\Phi_0=hc/2e$ and 
$K(\vec{x}-\vec{x}^\prime,h)\propto{}h/(\vert\vec{x}-\vec{x}^\prime%
\vert^2+h^2)^{3/2}$ for $\vert\vec{x}-\vec{x}^\prime\vert^2+h^2\gg{}%
a^2$ \cite{lee}, where $a$ is the lattice constant of the array.  The total 
flux, $\Phi(t)$, detected by the SQUID is 
$\Phi(t)=\int{}d^2xB_\perp(\vec{x},h=d;t)$, where the integral is performed 
over the effective area of the SQUID, $\ell_{\rm{}eff}^2=\ell_i\ell_o$. 
In fact, magnetic field lines produced by vortices near the SQUID are 
distorted by the presence of the superconducting washer, modifying the 
form of $K$. However, we do not expect this modification to change the 
scaling dimension of $K$ from two (that is, if we replace 
$\vec{x},\vec{x}^\prime,d$ by 
$\vec{x}/\xi,\vec{x}^\prime/\xi,d/\xi$, we still expect 
$K(\vert\vec{x}-\vec{x}^\prime\vert,d)$ to become 
$\xi^{-2}K\biglb((\vec{x}-\vec{x}^\prime)/\xi,d/\xi\bigrb)$ so that our 
scaling arguments remain valid.  Thus the flux-flux correlation 
function, $C_\Phi(t)\equiv\langle\Phi(t)\Phi(0)\rangle$, is 
\begin{eqnarray} 
C_\Phi(t)=&&\Phi_0\int{}d^2xd^2yd^2x^\prime{}d^2y^\prime{} 
K(\vec{x}-\vec{x}^\prime,d)\nonumber\\ 
&&\times{}K(\vec{y}-\vec{y}^\prime,d) 
\langle\rho_v(\vec{x}^\prime,t)\rho_v(\vec{y}^\prime,0)\rangle. 
\label{correlation} 
\end{eqnarray} 
Here, $\langle\ \rangle$ denotes the thermodynamic average, and the 
unprimed integrals are taken over the effective area of the SQUID.  Next, 
we assume a scaling ansatz for the vortex density-density correlation 
function: 
\begin{equation} 
\langle\rho_v(\vec{x}^\prime,t)\rho_v(\vec{y}^\prime,0) 
\rangle=\xi^{-4}F_1(t/\tau,\vert\vec{x}^\prime-\vec{y}^\prime\vert/\xi,%
L/\xi). 
\label{rhorho} 
\end{equation} 
Here we have used the fact that the scaling dimension of the vortex 
density is two at the KTB transition (that is, $\rho_v\propto\xi^{-2})$.  In 
Eq.\ (\ref{rhorho}), $F_1$ is a scaling function and $L$ is the smaller 
dimension of the array.  For Josephson junction arrays, 
\begin{mathletters} 
\label{csiandtprime} 
\begin{eqnarray} 
\xi=\xi_0\exp\Bigl(b/\sqrt{T^\prime-T_{\rm{}KTB}^\prime}\Bigr),\\ 
\label{csi} 
T^\prime\equiv{}k_BT/E_J(T)=2ek_BT/\hbar{}i_c(T), 
\label{tprime} 
\end{eqnarray} 
\end{mathletters} 
and $E_J(T)$ is the Josephson coupling energy per junction \cite{lobb83}. 
If we substitute the scaling ansatz, Eq.\ (\ref{rhorho}), into Eq.\ 
(\ref{correlation}) and perform proper rescaling of the integration 
variables, one can show that 
\begin{equation} 
C_\Phi(t)=\Phi_0^2F_2(t/\tau,d/\xi,L/\xi,\ell_{\rm{}eff}/\xi), 
\label{scaledcorr} 
\end{equation} 
where $F_2$ is a new scaling function.  The noise spectrum, $S_\Phi(f)$, 
is defined as $S_\Phi(f)=\int{}dt\exp(i2\pi{}ft)C_\Phi(t)$.  Substituting 
Eq.\ (\ref{scaledcorr}) in this expression, we find 
\begin{equation} 
fS_\Phi(T)=\Phi_0^2F(f/f_\xi,d/\xi,L/\xi,\ell_{\rm{}eff}/\xi). 
\label{noisegeneral} 
\end{equation} 
Here $F$ is another scaling function, the form of which we determine 
from a subsequent data collapse. 
 
\begin{figure}
\begin{center}
\epsfig{file=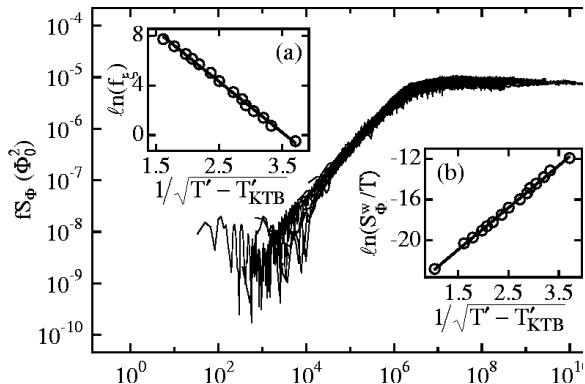,width=3in}
\end{center}
\caption{$fS_\Phi(f)$ versus $f/\tau_0f_\xi$ for flux noise shown in Fig.\ 
\ref{rawdata}; $T_{\rm{}KTB}=1.63\text{ K}, bz=4$.  Insets show 
$\ln(f_\xi)$ versus $1/\protect\sqrt{T^\prime-T_{\rm{}KTB}^\prime}$ with 
line of slope $-4.10$ and $\ln(S_\Phi^w/T)$ versus 
$1/\protect\sqrt{T^\prime-T_{\rm{}KTB}^\prime}$ with line of slope 4.13.}
\label{collapse}
\end{figure}

If we ignore the dependence on $d/\xi,L/\xi,\text{and }\ell_{\rm{}eff}/\xi$, 
Eq.\ (\ref{noisegeneral}) predicts 
\begin{equation} 
fS_\Phi(T)=\Phi_0^2F\Biglb((f/f_0)\exp 
\Bigl(bz/\sqrt{T^\prime-T_{\rm{}KTB}^\prime}\Bigr)\Bigrb), 
\label{noisespecific} 
\end{equation} 
where we have explicitly put in the temperature dependence of $f_\xi$.  In 
Eq.\ (\ref{noisespecific}), $T_{\rm{}KTB}$ and $bz$ are unknown.  To 
determine $T^\prime$, we fit $i_c(T)$ between $T=1.52\text{ K and } 
T=1.93\text{ K}$ using the expected temperature dependence 
$i_c(T)=i_c(0)\exp(-\alpha\sqrt{T})$ \cite{degennes64}, finding 
$i_c(0)=0.133\text{ A}$ and $\alpha=9.14\text{ K}^{-1/2}$, and extrapolate 
to higher temperatures.  Equation (\ref{noisespecific}) predicts that by 
choosing the correct values for $T_{\rm{}KTB}$ and $bz$ and plotting 
$fS_\Phi(f)$ versus the scaled frequency 
$f/\tau_0f_\xi=f\exp\bigl(bz/\sqrt{T^\prime-T_{\rm{}KTB}^\prime}\bigr)$, 
we should obtain a collapse of the raw data onto a single curve.  Our 
procedure for data collapsing is to make iterative changes in $bz$ and 
$T_{\rm{}KTB}$ until the best data collapse is obtained.  The final result is 
shown in Fig.\ \ref{collapse}, where the fitting parameters were $bz=4$ 
and $T_{\rm{}KTB}=1.63\text{ K}$.  The quality of the data collapse is 
strongly affected by the choice of $T_{\rm{}KTB}$ but relatively weakly by 
the choice of $bz$.  One can compensate an increase (decrease) of 
$T_{\rm{}KTB}$ of a few mK by a decrease (increase) in $bz$ of about 
10\%.  However, changes in $T_{\rm{}KTB}$ beyond a few mK or 
changes in $bz$ of more than about 10\% always result in a lower quality 
data collapse.  We can obtain a more accurate estimate of $bz$ through 
the temperature dependence of $f_\xi(T)$.  Using 
Eqs.\ (\ref{csiandtprime}) and 
$f_\xi(T)=f_0[\xi(T)/\xi_0]^{-z}$, we expect a straight line of slope $-bz$ 
on a plot of $\ln(f_\xi)$ versus 
$1/\sqrt{T^\prime-T_{\rm{}KTB}^\prime}$.  Inset (a) to Fig.\ 
\ref{collapse}, using $T_{\rm{}KTB}=1.63\text{ K}$, shows the resultant 
straight line which yields $bz=4.10\pm0.04$ and 
$f_0=(2.1\pm0.3)\times10^6\text{ Hz}$. 
 
We note that although $f_0$ is in the megahertz frequency range, the 
exponential dependence of $f_\xi$ on $T^\prime-T_{\rm{}KTB}^\prime$ 
implies that $f_\xi$ is reduced to the order of 1 Hz even when $T$ is 
about 0.2 K above $T_{\rm{}KTB}$.  This sensitivity is illustrated by the 
fact that the apparent transition temperature, 
$T_{\rm{}KTB}\text{(47 Hz)}$, associated with the sharp rise in $dV/dI$ 
measured at 47 Hz (inset, Fig.\ \ref{rawdata}) is substantially higher than 
the static value $T_{\rm{}KTB}=1.63\text{ K}$ inferred from our noise 
data.  However, these values are quite consistent since 
$T_{\rm{}KTB}\text{(47 Hz)}$ is the temperature at which 
$f_\xi=47\text{ Hz}$.  Using the above values of 
$T_{\rm{}KTB},f_0,\text{ and }bz$, we predict 
$T_{\rm{}KTB}\text{(47 Hz)}=1.94\text{ K}$, in good agreement with the 
$dV/dI$ data (inset, Fig.\ \ref{rawdata}). 
 
In addition to finding $bz$, we can extract the critical exponent $z$ in the 
following manner.  At low frequencies, a two-dimensional sample with a 
linear conductivity $\sigma$ produces white noise with 
spectral density $S_\Phi^w\propto{}k_BT\sigma$.  In the KTB regime for 
$f\ll{}f_\xi,\sigma$ is proportional to $\langle\rho_v\rangle^{-1}=\xi^2$, 
reflecting the intuitive idea that the low frequency conductivity is inversely 
proportional to the
\noindent 
density of free vortices.  Therefore, for $f\ll{}f_\xi$, 
\begin{equation} 
S_\Phi^w\propto{}k_BT\xi_0^2\exp\Bigl(2b/\sqrt{T^\prime- 
T_{\rm{}KTB}^\prime}\Bigr). 
\label{swhite} 
\end{equation} 
To test Eq.\ (\ref{swhite}), we define $S_\Phi^w$ as the horizontal line 
drawn through the low frequency data at each temperature, as shown for 
$T=1.978\text{ K}$ in Fig.\ \ref{rawdata}, and plot $\ln(S_\Phi^w/T)$ 
versus $1/\sqrt{T^\prime-T_{\rm{}KTB}^\prime}$.  Inset (b) to Fig.\ 
\ref{collapse}, using $T_{\rm{}KTB}=1.63\text{ K}$, shows the 
resultant straight line which yields $2b=4.13\pm0.04$.  Combining this 
result with the value $bz=4.10\pm0.04$ from the temperature 
dependence of $f_\xi$ yields the dynamic exponent $z=1.98\pm0.03$. 
 
In summary, we have used measurements of magnetic flux noise to study 
the equilibrium KTB transition in an array of overdamped Josephson 
junctions.  We emphasize that confirmation of the critical dynamics of the 
KTB transition requires measurement of frequency and temperature 
dependent properties such as those presented above.  Using  $bz=4$ and 
$T_{\rm{}KTB}=1.63\text{ K}$, we have shown that the data collapse over 
more than five decades in frequency, confirming the predictions of 
dynamic scaling.  In addition, from the temperature dependences of 
$f_\xi$ and $S_\Phi^w$, we have extracted the dynamic exponent 
$z=1.98\pm0.03$.  Our experimental finding $S_\Phi(f)\propto{}1/f$ in 
the critical frequency regime is inconsistent with theoretical predictions 
based on time-dependent Ginzburg-Landau (TDGL) dynamics with the 
classical two-dimensional XY model \cite{lee,houlrik94}.  The failure of 
this model is surprising since TDGL dynamics does predict 
$z=2$, in agreement with our extracted value.  It is unclear to us why 
$S_\Phi(f,T)$ is independent of $d/\xi,L/\xi,\text{and }\ell_{\rm{}eff}/\xi$. 
A possible cause may be that the data in Fig.\ \ref{rawdata} are 
in the regime $\xi(T)<\ell_{\rm{}eff}<L$.  Our evidence for this restriction 
is that at temperatures between $T_{\rm KTB}$ and $1.825$ K (the lowest 
temperature for which data are shown in Fig.\ \ref{rawdata}), we observe 
discrete jumps in the flux threading the SQUID.  We interpret this 
behavior as the motion of a single vortex under the SQUID, implying that 
$\xi(T)>\ell_{\rm{}eff}$, and conclude that $\xi(T)<\ell_{\rm{}eff}$ 
for the temperatures referred to in Fig.\ \ref{rawdata}.  
If we assume $\xi(T=1.825\text{ K})<\ell_{\rm{}eff}$, we deduce 
$\xi_0<0.2\ \mu\text{m}$ [Eqs.\ (\ref{csiandtprime})], a value that is 
considerably smaller than the commonly accepted lore $\xi_0=a$. 
Setting $\xi_0<0.2\ \mu\text{m}$, we deduce $\xi<2\ \mu\text{m}$ 
for the highest temperature ($2.379$ K) data shown in Fig.\ \ref{rawdata}.  
The fact that scaling persists to this temperaure where the correlation 
length is smaller than the lattice constant $a$ is somewhat disturbing. 
Further work is needed to resolve this issue, as well as the lack of 
dependence of $F$ on $d/\xi,L/\xi,\text{and }\ell_{\rm{}eff}/\xi$, and the 
$1/f$ behavior of the flux noise in the critical regime. 
 
In closing, we note that we have carried out similar measurements on two 
other arrays, one of square geometry and the other of triangular 
geometry, using the SQUID described above as well as two other SQUIDS 
of different geometry.  When the apparatus was cooled in a magnetic 
field ($<1\ \mu$T), the data were similar to those described above with 
minor differences in the high frequency behavior.  When a small field was 
applied, the noise data exhibited very different behavior.  These 
additional measurements will be described elsewhere. 
 
We thank Roger Koch for helpful discussions and Chris Lobb for sharing 
his constructive criticisms and knowledge of Josephson junction arrays 
with us.  This work was supported by the Director, Office of Energy 
Research, Office of Basic Energy Sciences, Materials Science Division of 
the U. S. Department of Energy under contract number 
DE-AC03-76SF00098. 

\end{document}